\documentclass[10pt,a4paper]{article}
\usepackage[T1]{fontenc}
\usepackage[intlimits]{amsmath} 
\usepackage{amsfonts}
\usepackage{graphicx} 
\def\tr{{\rm Tr}}
\def\Z{{\mathbb Z}} 
\def\C{{\mathbb C}}
\def\R{{\mathbb R}}
\def\SLn{SL(n,\mathbb R)} 

\def\SLketto{SL(2,\mathbb R)} 
\def\slketto{sl(2,\mathbb R)}
\def\SLharom{SL(3,\mathbb R)} 

\def\gotg{{\mathfrak g}}
\def\nyil{\, \longrightarrow \,}
\def\tnyil{\, \longmapsto \,}
\def\bea{\begin{eqnarray}}
\def\eea{\end{eqnarray}}
\def\nn{\nonumber}

\begin{document}

\title{Geometry of $W$-algebras from the affine \\ Lie algebra point of view}
\author{Z. Bajnok$^{1}$ and D. N\'ogr\'adi$^{1,2}$}
\date{December, 2000}
\maketitle

\vspace{-5.5cm}
\begin{flushright}
ITP-Budapest 562
\end{flushright}
\vspace{3.8cm}

\begin{center}
$^{1}$\emph{Institute for Theoretical Physics, E\"otv\"os University \\
 Budapest 1117, P\'azm\'any P\'eter s\'et\'any 1/A, Hungary}
\end{center}

\begin{center}
$^{2}$\emph{Institute Lorentz, University of Leiden \\
 P.O. Box 9506, NL-2300 RA Leiden, The Netherlands}
\end{center}

\vspace{0.5cm}

\begin{abstract}
To classify the classical field theories with $W$-symmetry one has to classify
the symplectic leaves of the corresponding $W$-algebra, which are the intersection
of the defining constraint and the coadjoint orbit of the affine Lie algebra if 
the $W$-algebra in question is obtained by reducing a WZNW model. The fields that
survive the reduction will obey non-linear Poisson bracket (or commutator) relations
in general. For example the Toda models are well-known
theories which possess such a non-linear $W$-symmetry and many features of these
models can only be understood if one investigates the reduction procedure.
In this paper we analyze the $\SLn$ case from which the so-called $W_n$-algebras
can be obtained. One advantage of the reduction viewpoint is that it gives a constructive
way to classify the symplectic leaves of the $W$-algebra
which we had done in the $n=2$ case which will
correspond to the coadjoint orbits of the Virasoro algebra and for
$n=3$ which case gives rise to the Zamolodchikov algebra. Our method in principle
is capable of constructing explicit representatives on each leaf. Another attractive
feature of this approach is the fact that the global nature of the $W$-transformations
can be explicitly described. The reduction method also enables one to determine
the classical highest weight (h. w.) states which are the stable minima of the energy on a $W$-leaf. These
are important as only to those leaves can a highest weight representation
space of the $W$-algebra be associated which contains a classical h. w. state.
\end{abstract}

\section{Introduction}
\label{intro}

$W$-algebras have attracted a great interest since their first appearance \cite{gelfand1}.
They are extensions of the Virasoro algebra by higher spin currents in general in such
a way that the Poisson bracket or commutator of the currents give non-linear terms
$$[W_i,W_j] = P_{ij}(W) \; ,$$
where $P_{ij}(W)$ is a polinomial of the currents.
Such extensions are relevant not only in the classification of two
dimensional conformal field theories but also in describing various statistical
physical models.

There are various ways to obtain the so-called $W_n$-algebras on
both quantum and classical level;
see \cite{zam, zam2} for the quantum field theoretical aspects where the degenerate
representations, the corresponding null vectors and quantum highest weights were constructed for
$W_3$ and was already pointed out how this algebra was related to $sl(3)$. Later it was shown in
\cite{todatheory} that the Toda models which carry the $W_n$-algebras as symmetry algebras
are Hamiltonian reductions of the Wess-Zumino-Novikov-Witten (WZNW) models. The
classical interpretation
of the $W_n$-algebras was discussed from a similar viewpoint also in \cite{origin} where
the conformal Ward-identities were also derived. In \cite{gervais, gervais2}
a different realization was proposed, the classical $W_n$-algebras were related to
certain embeddings of two dimensional manifolds into $2n$ dimensional K\"ahler manifolds which
construction makes global considerations possible and quite explicit. A link between the quantum
and classical levels was established in \cite{affine} and \cite{a2toda}. The
classical realization was studied geometrically in \cite{affine}
where the equation of motion was originated
from immersions of affine surfaces of constant mean curvature in affine space and was shown to be
the classical $(c \rightarrow - \infty)$ limit of Zamolodchikov's null vector equation.
In \cite{a2toda} the null vector equation was obtained as quantization of the classical equation
of motion for $W_3$.

Once the classical $W$-algebra is given - no matter where it originates from, we have seen
there are various ways to arrive at them - one can ask about its symplectic leaves
and classical highest weight states. This is the subject of the present paper. We
will follow the gauged WZNW model construction as this approach is the most useful
one for our purposes as it will become clear, but we would like to stress
that other classical interpretations of $W$-algebras mentioned above are equally good and the
problems we address are independent of how the classical algebras were arrived at.

A ``phase space'' associated to a Poisson manifold is one of its symplectic leaves.
For the exploration of the leaves we will use the well-known classical result stating
roughly that if second class constraints are imposed on a Poisson manifold $P$ resulting
in a submanifold $C \subset P$ (which is also Poisson in a canonical way) then
the symplectic leaves $M \subset C$ of $C$ coincide with
the symplectic leaves $S \subset P$ of $P$ intersected by $C$.

In our case the Poisson manifold $P$, as we have Kac-Moody symmetry, is
the Kac-Moody algebra with its canonical Poisson structure that comes from
the Lie algebra structure. Its symplectic leaves $S$ are the connected components of
its coadjoint orbits and the $W$-algebra lives on the constraint submanifold $C$ determined
by the Toda-type constraints.
The statement above means that the symplectic leaves of the $W$-algebra are 
nothing else then the connected components of the intersection of the constraint and
the Kac-Moody coadjoint orbits. This implies that to classify
the classical theories with $W$-symmetry, which is the same as
to classify the symplectic leaves of the $W$-algebra, one has to classify
the connected components of the intersection of the Kac-Moody coadjoint orbits and 
the constraint hypersurface.

Exploring the symplectic leaves in this way enables one in principle to
give explicit representatives on each $W$-leaf which was carried out in the
$n=2$ case but became technically quite cumbersome already if $n=3$.

The issue of the classical h. w. state i.e. a stable minimum of the energy can
also be handled by the reduction as the energy functional and its first and second
variation on a $W$-leaf will first be considered on the Kac-Moody coadjoint
orbit and then restricted to the intersection. We will determine which orbits
contain a classical h. w. state which will indicate that these can give rise to 
a highest weight representation space of the $W$-algebra. 

Yet another advantage of the reduction viewpoint is that we can implement the
$W$-transformations as gauge preserving Kac-Moody transformations. These
were given in \cite{todatheory} on the Lie algebra level but our analysis
makes it possible to explicitly describe them on the group level. These are
important for understanding the global nature of $W$-transformations.

Our motivation for analyzing the classical geometry of $W$-algebras is at least twofold.
First consider quantization.
The quantization of $W$-algebras started by a free field construction \cite{lukyanov1},
then a BRST method \cite{kac} was adopted to produce their quantum counterparts.
None of the approaches above however, relied on the geometry of $W$-algebras
loosing useful information in this way. For example one possible quantization procedure
would be the geometric quantization of the symplectic leaves for which it is
essential to explore the underlying geometry.

Our second motivation comes from a more mathematical context. There is a completely
different viewpoint of the classification problem which
has its origin in the field of integrable models. In this approach the classification
of symplectic leaves is related to the homotopy classes of certain non-degenerate
curves on projective space or sphere which problem is still unsolved in the general
case \cite{symplectic, shapiro}. We will try to make the relation of the two approaches clear
but will not elaborate much on this direction. Note that this formulation is close in
spirit to a third approach which addresses the problem throught the second Gel'fand-Dikii
bracket \cite{gelfand2}.

In section \ref{wznw} we will summarize the well-known setting of the WZNW models
concentrating on the symmetry properties. Section \ref{hurok} will be a summary
of those features of the geometry of loop groups and their central extensions which
are necessary for our considerations. This section will closely follow
\cite{loop}. After these rather general remarks we will set the stage for the reduction procedure
in the case of the $\SLn$ WZNW model in section \ref{sln} for which the main reference
is \cite{todatheory}. The detailed analysis
will be done in the two simplest cases; the $\SLketto$ case will amount to the classification
of the coadjoint orbits of the Virasoro algebra and the well-known result will be
recovered \cite{symplectic, coadjoint1, coadjoint2, unitary}. The $\SLharom$
case in section \ref{sl3w} will produce the
simplest $W$-algebra, the so-called Zamolodchikov algebra and we will mainly focus
on the behaviour of the energy functional on the $W$-leaves rather then giving
explicit representatives on each of them. We will summerize our and comment 
on other results and open questions in section \ref{conc}.

\section{WZNW models}
\label{wznw}

The usual action of the WZNW model on a connected, non-compact, simple Lie group is the
following:
$$S = \frac{k}{8 \pi} \int_{\Sigma} d^2 x \, \eta^{\mu \nu} \tr (g^{-1}
\partial_{\mu} g)(g^{-1} \partial_{\nu} g) - \frac{k}{12 \pi} \int_B
\tr (g^{-1} d g)^3 \; ,$$
where $k$ is the coupling constant, $\Sigma=S^1 \times
\R $ is the $1+1$ dimensional Minkowski space, $B$ is a $3$-manifold
whose boundary is $\Sigma$, $\eta^{\mu \nu}$ is the Lorentzian metric on $\Sigma$,
and $\; g : \Sigma \nyil G$ is the dynamical field. If a certain element
in $H^3(G)$ is not zero then $k$ has to be an integer otherwise the action or
rather the transition amplitudes are ill defined \cite{gawedzki}. 

The equations of motion look in terms of the currents in light-cone coordinates as
$$\partial_{-} J_{+} = 0 \; , \quad \partial_{+} J_{-} = 0 \; ,$$
where $J_{+}, J_{-} : \Sigma \nyil \gotg$ are the Lie algebra valued currents,
$$J_{+}=k \partial_{+} g g^{-1}, \quad J_{-}=-k g^{-1} \partial_{-} g \; ,$$
and $x_{\pm} = t \pm x$ are the light-cone coordinates on $\Sigma$.
The solutions are simply that $J_{+}$ depends on $x_{+}$, $J_{-}$ on $x_{-}$ only,
which means they are periodic functions of one variable. This implies that the solution
in the language of the $g$ field is
$$g(x_+,x_-)=g_+(x_+) g_-(x_-)$$
where $g_+$ and $g_-$ are quasiperiodic group valued fields which
means $g_+(x_+ + 2\pi)=g_+(x_+) M$ and that $g_-(x_- - 2\pi)=M^{-1} g_-(x_-)$ for some
$M \in G$ monodromy.

The action of an $x_{+}$ dependent chiral $h(x_+)$ Kac-Moody transformation on the fields are:
\bea
\label{trafo}
g & \tnyil & h g h(0)^{-1} \nn \\
J_{+} & \tnyil & h J_{+} h^{-1} + k \partial_{+} h \, h^{-1} \; ,
\eea
and analogous formulas hold for $J_{-}$. These chiral transformations
are symmetries of the action and we will see that these are nothing but the coadjoint
action of the centrally extended loop group, and to make the identification
more transparent $h(0)^{-1}$ is defined into the transformation of $g$. This
modification is actually a combination of the left local,
$ g(x_+, x_-) \tnyil h(x_+) g(x_+, x_-) $, and the right global,
$ g(x_+, x_-) \tnyil g(x_+, x_-) h(0)^{-1} $, 
transformations which are both symmetries of the action. The conservation
laws that correspond to these symmetries coincide with the equations of motion.

From this point forward we will only deal with the ``$+$'' chirality and $J$, $g$ and $x$ will
mean $J_{+}$, $g_+$ and $x_{+}$. So $g$ and $J$ are functions of one variable, $J$ is periodic 
$g$ is quasiperiodic only.

As the $J$ current generates the (\ref{trafo}) transformations which will turn out to be
the coadjoint action of the centrally extended loop group the Poisson brackets of 
its components coincide of course with the commutation relations of the Kac-Moody algebra.
Let $\{t^a\}$ be a basis of $\gotg$, then $J(x)=J_a(x) t^a$ and
$$\{J_{a}(x),J_{b}(y)\}=f_{ab}^c J_{c}(x) \delta (x-y) + k K_{ab} \delta^{\prime}(x-y) \; ,$$
where $\delta(x)$ is the $2\pi$ periodic $\delta$-function, $f_{ab}^c$
are the Lie algebra structural constants and $K_{ab}$ is the Killing form.

\section{Loop groups and their central extensions and coadjoint orbits}
\label{hurok}

Let $G$ be a connected and simple Lie group. The loop group $LG$, associated to $G$ is
the group of smooth $S^1 \nyil G$ mappings with pointwise multiplication \cite{loop}. The
following observation will be important: if $G$ is not simply connected then $LG$ is not
connected as the connected components of $LG$ are labeled by the fundamental group of $G$.

Let us parametrize the circle with $x$ ranging from $0$ to $2 \pi$. The map
$h \tnyil (h(0), h \, h(0)^{-1})$ is a diffeomorphism between $LG$ and
$G \times \Omega G$ where $\Omega G$ is the group of such loops that start at the
identity. On $G \times \Omega G$ the group multiplication is a semi-direct product:
$$(A, \gamma)(B, \eta) = (A B, \, \gamma \,
A \eta A^{-1}), \quad A, B \in G, \quad \gamma, \eta \in \Omega G \; ,$$
so $\gamma(0) = \eta(0) = 1$. The Lie algebra of $LG$, $L\gotg$, is the space of smooth
$S^1 \nyil \gotg$ mappings, the Lie algebra of $\Omega G$, $\Omega \gotg$, is the space of
such loops that start at the origin.

The Kac-Moody algebras are the central extensions of $L\gotg$, they are denoted by $\hat \gotg$.
Since $G$ is simple the extension by $\R$ is unique up to a scalar factor and the 
corresponding cocycle is $c(J_1, J_2) = \langle J_1^{\prime}, J_2 \rangle$, where
\bea
\label{nemtom}
\langle J_1, J_2 \rangle = \int_{0}^{2 \pi} \tr \, J_{1}(x)J_{2}(x) \frac{dx}{2 \pi} \;
\eea
is a non-degenerate form on $L\gotg$. Note that the cocycle
$c:L \gotg \times L \gotg \nyil \R$ is 
degenerate, its kernel consists of precisely the constant elements but if it is
restricted to $\Omega \gotg \times \Omega \gotg$ it will turn non-degenerate as 
the constant elements in $\Omega \gotg$ are zero.

While considering the coadjoint orbits of the centrally extended loop groups it is clearly enough
to deal with the action of $LG$ and $L\gotg$ on $\hat \gotg$ as we have a central
extension at hand. Note that we have not touched the delicate issue of central extensions
on the group level and we will continue to ignore this subtlety. It is easy to check
that the adjoint action of $K \in L\gotg$ and $h \in LG$ on $\hat \gotg$ is
$$K \cdot (J,k) = ([K,J],c(K, J)) \; ,$$
$$h \cdot (J,k) = (h J h^{-1}, k - \langle h^{-1} h^{\prime}, J \rangle) \; .$$
The coadjoint action will be considered also on $\hat \gotg$ after identifying the
dual of $L\gotg$ with $L\gotg$ by the (\ref{nemtom}) scalar product and 
turns out to be
$$h \cdot (J,k) = (h J h^{-1}+k h^{\prime} h^{-1},k) \; ,$$
where $\cdot$ now of course means the coadjoint action. In the first component we have
the symmetry transformation (\ref{trafo}) of the WZNW model and the central term is invariant, so
we will think of the coadjoint action of the centrally extended loop group as an action
of $LG$ on $L\gotg$ with a given $k$ fixed. The corresponding orbit will be referred
to as the Kac-Moody coadjoint orbit or simply as the coadjoint orbit.

Let us consider the quasiperiodic group valued field of the WZNW model with $M \in G$
monodromy for analyzing the coadjoint orbits. The association of $J$ to $g$ via the
$J=k g^{\prime} g^{-1}$ definition is one-to-one between the space of quasiperiodic group
valued fields and $L\gotg$ if we add the $g(0)=1$ initial condition which will mean that
$M=g(2\pi)$. Note that this initial condition is the reason for including the constant
term in (\ref{trafo}). To a given $J(x)$ corresponds the path ordered
$g(x)={\rm P} \exp (\int_0^x J(y) dy)$. If the transformation of $J$ is $J \tnyil h
J h^{-1} + k h^{\prime} h^{-1}$ then the associated $g$ transforms as $g \tnyil h g
h(0)^{-1}$ which are the symmetry transformations of the WZNW model. The monodromy also
changes to $M \tnyil h(0) M h(0)^{-1}$. This means that if two $J$s are on the same orbit
then the corresponding monodromies are conjugated in $G$. The other way around take
two elements in $L\gotg$, $J_1$ and $J_2$ and let the corresponding
quasiperiodic fields and monodromies be $g_1, g_2$ and $M_1$ and $M_2$. Let us assume that 
the monodromies are conjugated by an $A \in G$, $M_2=A M_1 A^{-1}$, then $g_2 A g_1^{-1}$
is periodic and takes $J_1$ to $J_2$ which means they are on the same orbit.

To summarize the above considerations there is an equivariant correspondence between
the conjugacy classes of $G$ and the Kac-Moody coadjoint orbits which fact greatly
simplifies the study of the latter. For example it is fairly straightforward to determine the 
isotropy subgroup, $H_J \subset LG$, of a given $J$ because if $h$ leaves $J$ fixed then
it must be true that $g = h g h(0)^{-1}$ and $h(0)$ commutes with the corresponding 
monodromy, which means that $h$ is such that $h = g A g^{-1}$ where $h(0)=A$ is any
element which commutes with $M$. This implies that the map $h \tnyil h(0)$ is an 
isomorphism from $H_J$ to the commutant subgroup of $M$ in $G$, hence we will think
of $H_J$ as a subgroup of $G$.

In general any orbit can be realized as a homogenous manifold which gives in our
case the realization ${\cal O}_J = LG / H_J = (G \times \Omega G) / H_J$ where ${\cal O}_J$
is the coadjoint orbit of $J$. As $H_J$ can be thought of as a subgroup of $G$ one
might suspect that the factorization only takes place in the first ($G$) component. To make 
this idea precise consider the map:
\bea
\label{nyalab}
\pi : {\cal O}_J & \nyil & G/H_J \nn \\
h \! \cdot \! J & \tnyil & [h(0)] \; ,
\eea
where $\cdot$ is the coadjoint action and $[h(0)]$ is the class of $h(0)$ in $G/H_J$.
This map is easily seen to be well defined. It is clear that $\pi^{-1}([A])$ is nothing
other then those elements in ${\cal O}_J$ whose monodromies are $A M A^{-1}$ and that on this
set $\Omega G$ acts freely as if the monodromy of $g$ is $A M A^{-1}$ and
$\gamma \in \Omega G$ then the monodromy of $\gamma g \gamma(0)^{-1}=\gamma g$ is the same.
Local trivializability can also be established. Summarizing, the coadjoint
orbits of Kac-Moody groups are principal
$\Omega G$-bundles over $G / H_J$, hence all the infinite dimensional subtleties were
isolated in the fiber and is the same for all orbit and the orbit specific features were
captured in the finite dimensional homogenous manifold $G / H_J$.

Let us elaborate a little more on the important special case when the monodromy $M$ has
a logarithm in $\gotg$ i.e. it is of the form $M=\exp(X)$ for some $X \in \gotg$. Having 
such an $M$ at hand an obvious choice for the quasiperiodic field with this monodromy is
$g(x)=M^{\frac{x}{2\pi}}$. To this $g$ corresponds the
constant current $J=\frac{1}{2\pi} X$ which shows that in this special case the orbit contains 
a constant representative which is not true in general for non-compact groups. Put it
the other way it is quite obvious that if an orbit contains a constant
representative then the corresponding monodromy has a logarithm. 

The isotropy subgroup of a constant $J=\frac{1}{2\pi} X$ consists of only those constant
elements in $LG$ which commute with $M$ which is the same as the isotropy subgroup of $X$, $H_X$,
with respect to the adjoint action of $G$ on $\gotg$. Recall that the (co)adjoint
orbit of $X$, ${\cal O}_X$, can be identified with $G / H_X$ which makes it clear that
in this case the ${\cal O}_J$ orbit is a principal $\Omega G$-bundle over 
${\cal O}_X$:
\bea
{\cal O}_J & \nyil & {\cal O}_X \nn \\
h \! \cdot \! J & \tnyil & h(0) X h(0)^{-1}\nn
\eea
and the map 
\bea
{\cal O}_J & \nyil & {\cal O}_X \times \Omega G \nn \\
h \! \cdot \! J & \tnyil & (h(0) X h(0)^{-1} ,h \, h(0)^{-1}) \nn
\eea
trivializes ${\cal O}_J$. Note that the (\ref{nyalab}) bundle can be trivialized in any case
over $G / H_J$ not only when $M$ possesses a logarithm but in general the trivializing map is
more complicated. 

The chiral energy in the WZNW model is of the usual form
\bea
\label{energy}
E(J)= \frac{1}{2} \langle J, J \rangle = \frac{1}{2} \int_0^{2\pi} \tr \, J^2(x) \frac{dx}{2\pi}
\eea
whose critical points on the orbit are the constant elements because if
$\delta J = [ K , J] + k K^{\prime} \,$ for a $K \in L\gotg$ then
\bea
\label{denergy}
\delta E = k \int_0^{2\pi} \tr J(x) K^{\prime}(x) \frac{dx}{2\pi}
\eea
which is obviously zero only for any $K$ if $J$ is constant. This can also be seen
by noting that $E(J)$ generates
the rigid rotations of the circle and the critical points should be invariant, hence they
must be constant. The critical points are
minima of the energy only if the second variation is positive semi-definite, which turns
out to be at a critical point evaluated on two elements $K_1, K_2 \in L\gotg$
\bea
\label{ddenergy}
\delta_2 \delta_1 E = k \int_0^{2\pi} \tr \left( J [K_1^{\prime}(x),K_2(x)] + k K_1^{\prime}(x) K_2^{\prime}(x) \right)  \frac{dx}{2\pi} \; .
\eea
which is symmetric in $K_1$ and $K_2$ for a constant $J$ as it should be. The positive
semi-definiteness of this quadratic form is equivalent to the positive semi-definiteness of the operator
$$-k^2 \left( \frac{d}{dx} \right)^2 + k {\rm{ad}} J \, \frac{d}{dx}$$
on L$\gotg$.

\section{$\SLn$ WZNW model and reduction}
\label{sln}

Systems with $W$-symmetry can be obtained from WZNW models based on maximally non-compact
Lie groups by imposing appropriate constraints. In general an $\slketto$ embedding
is needed in the Lie algebra in order to define the constraints but we will only
deal with the principal embedding in the $\SLn$ case from which generalization to
other groups and embeddings is straightforward.
From now on $J / k$ is denoted by $J$ (assuming that $k \neq 0$)
and $k$ will not appear explicitly in the formulas. The constraint is the fixing
of certain components of the current:
$$J_{\alpha}(x) =\, \, \left\{ \begin{array}{l}
1 \; \textrm{ if } \; \alpha \in \Delta _{-} \\
0 \; \textrm{ if } \; \alpha \in \Phi_{-} \setminus \Delta_{-}
\end{array} \right. $$
if the Cartan basis is chosen in $\gotg$ and where
$\Phi_-$ is the set of negative roots and $\Delta_-$ is the set of simple negative roots. 

These constraints mean that the form of $J$ is restricted in the following way:
$$J= \left( \begin{array}{ccccccc} * & * & . & . & . & . & * \\ 1 & * & * & &
& & . \\ 0 & 1 & * & * & & & . \\ . & 0 & 1 & & & & . \\ . & & & . & &
& . \\ . & & & & . & * & . \\ 0 & . & . & . & 0 & 1 & * 
\end{array} \right) \; .$$

It is easy to show that these
are first class constraints in the sense of Dirac and generate gauge transformations which are
precisely the transformations generated by the components of the current corresponding to
the positive roots. One possible gauge fixing is the Wronsky gauge in
which the current takes the form
\bea
\label{fix}
J= \left( \begin{array}{ccccccc} 0 & W_{2} & W_{3} & W_{4} & . & . & W_{n} \\
1 & 0 & 0 & . & . & . & 0 \\ 0 & 1 & 0 & & & & . \\ . & 0 & 1 & & & &
. \\ . & & & . & & & . \\ . & & & & . & 0 & . \\ 0 & . & . & . & 0 & 1
& 0 
\end{array} \right) \; .
\eea
One can think of the above form as the constrained form of the current by
second class constraints and can calculate the Poisson bracket of the reduced
$W_2, \ldots , W_n$ fields three different ways. As gauge invariant quantities one
can use the original Poisson bracket, or as fields left from second class constraints
one can adopt the Dirac formalism. The third possibility is based on the fact (more
or less definition) that the Poisson bracket taken with a $W_m$ field is the same
as the infinitezimal $W$-transformation generated by it and the brackets can be read off
from the transformation properties. The bracket of the $W_2=L$ energy-momentum tensor with
itself gives the Virasoro algebra of course and suitable combinations of the other $W_m$ fields transform as
primary fields of weight $m$ under conformal transformations. The bracket of two $W_m$
fields ($m \neq 2$) gives non-linear expressions in the others. This
Poisson algebra is called the $W_n$-algebra.

The energy-momentum tensor generates the conform transformations out of which the rigid
rotations are generated by the chiral energy
$E=L_0= \int_{0}^{2 \pi} L(x) \frac{dx}{2 \pi}$, hence the critical points of the
energy on a $W$-leaf
are those points which are invariant under rigid rotations i.e. all $W_m$ fields must be 
constant which means that this property has survived the reduction. 

The constraint on the current gives also restrictions on the $g$ field. For the current
to have the gauge fixed (\ref{fix}) form $g$ must take the form
\bea
\label{gfix}
g= \left( \begin{array}{c} \; \; \quad \quad {\textrm  {\boldmath $\psi$}}^{(n-1)} \; \; \quad \quad \\ \hrulefill \\
\vdots  \\ \hrulefill \\ {\textrm {\boldmath $\psi$}}^{\prime}\\
\hrulefill \\ {\textrm {\boldmath $\psi$}}
\end{array} \right) \; ,
\eea
where \mbox{\boldmath $\psi$} $: \R \nyil \R^n$ is a quasiperiodic row vector with $M$
monodromy and \mbox{\boldmath $\psi$}$^{(n-1)}$ is the $(n-1)^{th}$ derivative. The $\det{g}=1$
and $g(0)=1$ conditions mean further restrictions on \mbox{\boldmath $\psi$} and it follows
from the $J = g^{\prime} g^{-1}$ relation that its components are the linearly independent
solutions of the 
\bea
\label{psidiff} 
\psi^{(n)} = \sum_{m=2}^{n-1} W_m \psi^{(n-m)}
\eea
$n^{th}$ order linear differential equation which is the field equation of the reduced
theory in one chirality. Note that in case $n=2$ equation (\ref{psidiff}) turns into
the Hill equation which was used in \cite{coadjoint1} for analyzing the coadjoint orbits
of the Virasoro algebra. Also note that if \mbox{\boldmath $\psi$} in (\ref{gfix}) is
multiplied by any periodic function $r(x)$ then $ \det{g} $ picks up a factor of $r^{n}(x)$,
which shows that the length of \mbox{\boldmath $\psi$} is irrelevant, it can be thought
of as a quasiperiodic curve in $\R P^{n-1}$ or $S^{n-1}$ and the $\det{g} \neq 0$
condition is a certain non-degeneracy condition which makes it clear that the symplectic
leaf classification problem is equivalent to the determination of homotopy types of
quasiperiodic non-degenerate curves on projective space or the sphere. This approach
is followed in \cite{symplectic, shapiro}.

The Kac-Moody transformations which preserve the gauge fixed form of the current or
equivalently the (\ref{gfix}) constrained form of $g$
are also symmetries of the reduced model, these are the $W$-transformations. To 
give an explicit description of them let us consider the constrained $g$ field
in the form of (\ref{gfix}). It is not difficult to show that this form is preserved
by an $h \in LG$ Kac-Moody transformation
\bea
\label{megorzo}
h= \left( \begin{array}{c}  \quad \qquad {\mathbf a_n}  \quad \qquad \\ \hrulefill \\ \vdots
\\ \hrulefill \\ {\mathbf a_2}\\ \hrulefill \\
{\mathbf a_1}
\end{array} \right) \; ,
\eea
only if the ${\mathbf a_m} : S^1 \nyil \R^n$ row vectors satisfy the ${\mathbf
a_{m+1}}= {\mathbf a_{m}^{\prime} + {\mathbf a_{m}}J}, \; \; m=1 \ldots n-1$ recursion
relation whose solution is
$${\mathbf a_{m+1}(x)}={\mathbf a_1(x)} \left( \frac{d}{dx}+J(x) \right) ^m \; ,$$
where differentiation acts on the left. If the current is not in the gauge fixed form 
then the expression (\ref{megorzo})
gives those transformations which bring it to the gauge. As a consequence of
the recursion relation any component of $h$ can be expressed in terms of ${\mathbf a_1}$,
the only restriction on it is the $\det{h}=1$ condition which makes it possible to express
the last component of ${\mathbf a_1}$ with the others on the infinitezimal (Lie algebra) level.
The next to the last component generates the conformal the others the $W$-transformations. 

Based on the above reduction procedure it is not hard to find the number of independent
vectorfields that leave all the $W_m$ fields fixed by the corresponding $W$-transformation.
Because if it leaves them fixed one can neglect the constraint hypersurface and the answer
is the dimension of the isotropy subgroup $H_J \subset G$ of $J$. Attacking this
problem by brut-force methods would involve the determination of the number of independent
solutions of complicated differential equations subject to periodic boundary conditions which
is in general quite cumbersome. This direct method was used in \cite{coadjoint2} for 
determining the possible dimensions in the case of the Virasoro algebra.

To list all classical models with such a $W$-symmetry one has to classify the 
symplectic leaves of the non-linear Poisson structure of the $W_m$ fields. These
can be obtained, just as in the finite dimensional cases \cite{nonlinear, finite},
as the intersection of the 
coadjoint orbit of the original symmetry algebra, which is of course the $\SLn$ Kac-Moody 
algebra, and the constraint hypersurface as it can be implemented by
second class constraints. Or more precisely as a symplectic
leaf is connected by definition, the connected
components of the intersection should be determined. 

\section{$\SLketto$ Kac-Moody and Virasoro algebra}
\label{virasoro}

The only component of the current that survives
the reduction is the $W_2=L$ energy-momentum tensor which means that applying our method
in this case will result in the classification of the coadjoint orbits of the Virasoro
algebra and should reproduce the well-known results
\cite{symplectic, coadjoint1, coadjoint2, unitary}.
In \cite{coadjoint1} the authors classified the Virasoro orbits through
the study of the Liouville (or Hill) equation, hence our method will be a constructive way
of producing the conformally inequivalent solutions of the Liouville (or Hill) equation.

Also important to note that since $\pi_1(\SLketto) = \Z$, which is given by the 
winding number of the second row of the $\SLketto$ matrix, every
Kac-Moody orbit consists of $\Z$ connected components. If a $J$ current is given on a
specific component of an orbit then to take it to other components of the same orbit
one can apply for example the 
$$T_n(x)= \left( \begin{array}{cc} \cos(nx) & -\sin(nx) \\ \sin(nx) & \cos(nx)
\end{array} \right)$$
Kac-Moody transformation which has winding number $n$, where $n$ is any integer.
As the symplectic leaves are connected one has to determine the connected
components of the intersection of the Kac-Moody orbit and the constraint hypersurface.
It will turn out that the constraint either intersects each component of the Kac-Moody 
orbit in one piece or it does not intersect it at all. This means that the Virasoro
orbits can be labelled by the conjugacy class of the monodromy matrix and by those
$n$s to which component belongs an intersection. The issue of the existence of the intersection
will be dealt with in the following way, if a current is given on a specific component of 
a Kac-Moody orbit we will try to transform it to the desired 
$ \left( \begin{array}{cc} 0 & L \\ 1 & 0 
\end{array} \right) $
form by a Kac-Moody transformation that has winding number zero in order to
stay on the same component. If this can be done the resulting $L$ will immediately
be a representative of the Virasoro orbit, if it can not be done then the component in
question has no intersection with the constraint hypersurface.

Before the details note one more qualitative remark: according
to section \ref{sln} the energy-momentum tensor can be left fixed by $1$ or $3$ generators only 
as the group $\SLketto$ has only $1$ and $3$ dimensional isotropy subgroups, so in the
realization Diff$(S^1) / H$ of the Virasoro orbits the dimension of $H$ can only be
$1$ or $3$ as it is well-known. In the detailed analysis we will give the vectorfields
for each representative of the Virasoro orbit which leaves it fixed under the coadjoint
Virasoro action.

After imposing the constraint and the $\det{g}=1$ condition only one free
function $\phi$ remains in $g$:
$$g= \left( \begin{array}{cc} \left( \frac{\sin{\phi}}{\sqrt{\phi^{\prime}}} \right)^{\prime} &
\left( \frac{\cos{\phi}}{\sqrt{\phi^{\prime}}} \right)^{\prime} \\
\frac{\sin{\phi}}{\sqrt{\phi^{\prime}}} & \frac{\cos{\phi}}{\sqrt{\phi^{\prime}}} 
\end{array} \right) \; ,$$
where of course $\phi^{\prime} > 0$. The $g(0)=1$ initial condition and quasiperiodicity
of $g$ implies that
\bea
\label{kvaziperiod}
\phi(0)=0, \quad & & e^{i \phi(2\pi)}=\frac{d+ic}{\sqrt{d^2+c^2}} \nn \\
\phi^{\prime}(0)=1, \quad & & \phi^{\prime}(2\pi)=\frac{1}{d^2+c^2} \\
\phi^{\prime \prime}(0)=0, \quad & & \phi^{\prime \prime}(2\pi)=-2\frac{cd+ab}{(d^2+c^2)^2} \; ,\nn
\eea
where $M=\left( \begin{array}{cc} a & b \\ c & d 
\end{array} \right)$ is the monodromy matrix. The condition on $\phi(2\pi)$ and
the fact that $\phi$ is monotonically increasing implies that one can find an integer $N \geq 0$
such that
\bea
\label{phin}
\phi(2\pi)=\alpha+2\pi N \; ,
\eea
where $e^{i \alpha}=\frac{d+ic}{\sqrt{d^2+c^2}}$ and $0 < \alpha \leq 2\pi$. It does not
make sense to talk about the winding number of $\phi$ as it is only quasiperiodic but
the $N \geq 0$ integer just defined describes its winding in a certain sense. More precisely
let $J_1$ and $J_2$ be two gauge fixed elements on the same Kac-Moody orbit but not necessarily
on the same component. Their monodromies are conjugated by an $A \in \SLketto$, 
$M_2=A M_1 A^{-1}$. Let $g_1$ and $g_2$ be the quasiperiodic fields associated to them and
since they are gauge fixed they can be expressed with a $\phi_1$ and $\phi_2$ functions
whose invariant are denoted by $N_1$ and $N_2$. Then $J_1$ and $J_2$ are on the same
component if and only if $N_1=N_2$. One has only to prove that the winding number of
$h=g_2 A g_1^{-1}: S^1 \nyil \SLketto$ which takes $J_1$ to $J_2$ is zero if
and only if $N_1=N_2$ but this
is not difficult. So the $N$ invariant describes on which Kac-Moody component is the $J$
relative to a prescribed representative.

The space of $\phi$ functions with fixed monodromy
and $N$ invariant, subject to the (\ref{kvaziperiod}) conditions is clearly connected so if
the monodromy is fixed the constraint hypersurface can intersect each component in maximum
one piece. Besides, if the monodromy is changed within its conjugacy class
then in those components where there was an intersection it will continue to be so, which
means that the constraint hypersurface intersects a Kac-Moody component in either one piece
or not at all. It also turns out from this analysis that since $N \geq 0$ only those
Kac-Moody components can have an intersection with the constraint that are ``above'' the
prescribed representative which fact will be transparent in each specific case in the
detailed analysis. 

In each gauge fixed current the total degrees of freedom is in the corresponding $\phi$. The
energy-momentum tensor can be expressed as:
$$L=- {\phi^{\prime}}^2+\frac{1}{2} S(\phi) \; ,$$
where $S(\phi) = \frac{3}{2} \frac{{\phi^{\prime \prime}}^2}{{\phi^{\prime}}^2}-
\frac{\phi^{\prime \prime \prime}}{\phi^{\prime}}$ is the Schwarzian derivative. We have 
brought any $L$ to the above form which is quite remarkable as if it would be true
that $\phi(x+2 \pi) = \phi(x) + 2 \pi$ then this formula would mean that any $L$ is
on the orbit of $ -1 $ which is of course not true because the complicated
(\ref{kvaziperiod}) formulas hold for $\phi(x+2\pi)$ instead of the previous one. 

To classify the Kac-Moody and Virasoro orbits one has to present a complete list
of conjugacy classes. There are four distinct types of these depending
on the isotropy subgroup $H$.

\begin{itemize}

\item Elliptic classes, those monodromies belong here for which $|{\tr M}| < 2$, 
$$H=\left\{ \left. \left( \begin{array}{cc} \cos(t) & -\sin(t) \\
\sin(t) & \cos(t)
 \end{array} \right) \right| \; \; t \in (0,2 \pi ] \; \right\} ,$$
representatives:
$$M= \left( \begin{array}{cc} \cos(2 \pi \omega) & -\sin(2 \pi \omega) \\ \sin(2
\pi \omega) & \cos(2 \pi \omega)
\end{array} \right), \; \omega \in (0,1) \; , \; \omega \neq \frac{1}{2} \; .$$

\item Hyperbolic classes, those monodromies belong here for which $|{\tr M}| > 2$,

$$H=\left\{ \left. \left( \begin{array}{cc}
t & 0 \\ 0 & t^{-1}
\end{array} \right) \right| \; \; t \in \R^* \; \right\} ,$$
representatives:
$$M= \pm \left( \begin{array}{cc} e^{2 \pi b} & 0 \\ 0 & e^{-2 \pi b}
\end{array} \right), \; b > 0 \; .$$

\item  Parabolic classes, those monodromies belong here for which $|{\tr M}| = 2$,
but $M \neq \pm 1$,
$$H=\left\{ \left. \left( \begin{array}{cc} \pm 1 & 0 \\ t & \pm 1
\end{array} \right) \right| \; \; t \in \R \; \right\} ,$$
representatives:
$$M= \pm \left( \begin{array}{cc} 1 & 0 \\ q & 1
\end{array} \right), \;  q=\pm 1 \; .$$

\item Exceptional classes, $H=\SLketto$, representatives: $M=\pm 1$, these can be obtained
as the $\omega \nyil 0$ and $\omega \nyil \frac{1}{2}$ limiting cases of the elliptic
cases but the isotropy subgroups are larger. 

\end{itemize}

According to section \ref{sln} for a given $M$ one has to find a $g$ then the corresponding
$J$ current has to be studied whether it can be transformed to the gauge fixed form or not.
The most straightforward case is when $M$ possesses a logarithm in $\slketto$.
When this occurs the simplest
choice is $g(x) = M^{\frac{x}{2 \pi}}$ to which corresponds the constant current
$J_0 = \frac{1}{2 \pi} \log{M}$. From the above listed representatives the elliptic monodromies
have logarithms and from the others in the ``+'' case $M$, in the ``$-$'' case $\; -M$ has
a logarithm and in the latter case a possible choice for $g$ is
$g(x) = T_{\frac{1}{2}}(x) (-M)^{\frac{x}{2 \pi}}$ to which corresponds the non-constant 
current
$$J_{\frac{1}{2}}=T_{\frac{1}{2}} J_0 T_{- \frac{1}{2}} + T_{\frac{1}{2}}^{\prime} T_
{- \frac{1}{2}} .$$
Remember that $T_{\frac{1}{2}}(x)$ is the matrix of rotation by $\frac{x}{2}$.

From $J_0$ and $J_{\frac{1}{2}}$ we will get a representative on each Kac-Moody
component by applying $T_n$ to $J_0$ and (except for the elliptic case) to $J_{\frac{1}{2}}$.
To simplify notation from now on let $n$ be a half-integer and 
$$J_0=\left( \begin{array}{cc} j_1 & j_2-j_3 \\ j_2+j_3 & -j_1  \end{array} \right)$$
the constant element. We obtain a representative on each component of the Kac-Moody
orbits by setting
\bea
\label{jn}
J_n = T_{n} J_0 T_{-n} + T_{n}^{\prime} T_{-n} =
\eea
$$
=(n+j_3) \left( \begin{array}{cc} 0 & -1 \\ 1 & 0  \end{array} \right) + 
j_1 \left( \begin{array}{cc} \cos(2nx) & \sin(2nx) \\
\sin(2nx) & -\cos(2nx) \end{array} \right) +
j_2 \left( \begin{array}{cc} -\sin(2nx) & \cos(2nx) \\
\cos(2nx) & \sin(2nx) \end{array} \right)
$$
which is periodic also for half-integer $n$ as it should be. 

To summarize, the $J_n$ in equation (\ref{jn}) for integer $n$ is a representative of the $n^{th}$ component of
the Kac-Moody orbit corresponding to $M=e^{2\pi J_0}$ monodromy, and for half-integer $n$
it is a representative of the $(n- \frac{1}{2})^{th}$ component of the Kac-Moody orbit 
corresponding to $M=-e^{2\pi J_0}$ monodromy. As a reference let $J_0$
be on the $0^{th}$ component.

To get the Virasoro representatives $J_n$ has to be transformed to the gauge fixed
form with an $h \in L\,\SLketto$ with zero winding number. Obviously the winding number of
$$h=\left( \begin{array}{cc} R^{-1} & (1 + \frac{1}{n} R^{-2})R^{\prime}
\\ 0 & R  \end{array} \right):S^1 \nyil \SLketto$$
is zero and is easily checked to bring $J_n$ to the desired form as it is in
the form of (\ref{megorzo}) (and has no
singularity when $n=0$), and where
$$R(x)=\frac{1}{\sqrt{n+j_3+j_1 \sin(2nx) + j_2 \cos(2nx)}} \; .$$
This is only true of course if
\bea
\label{nagyobb}
n+j_3+j_1 \sin(2nx) + j_2 \cos(2nx) > 0 \; ,
\eea
which has to be checked in each case. If this inequality holds the Virasoro
representative turns out to be after straightforward calculations
$$L=C+2n(n^2+C)R^2+3n^2(C-n^2-2n j_3)R^4 \;,$$
where $C=\frac{1}{2} \tr (J_0)^2 = j_1^2+j_2^2-j_3^2$. This is a surprisingly dense form if we take into account
that it contains every possible case and besides it looks very similar to the
representatives given in \cite{coadjoint1} obtained by very different methods.
If the (\ref{nagyobb}) inequality does not hold then conjugating $J_0$ by a
constant element might change it in such a way that with the new $j_1$, $j_2$ and $j_3$
the inequality holds. This corresponds to choosing a different representative in
the conjugacy class of the monodromy. As we will see this will
occur in the hyperbolic case. In those
cases when this can not be achieved it is quite easy to show through the analysis of 
the $\det h=1$ equation, where $h$ is of the (\ref{megorzo}) form, that there is
no transformation that brings $J_n$ to the gauge
fixed form which means that the corresponding Kac-Moody component has no
intersection with the constraint hypersurface.

For the sake of completeness we will give the vectorfields for each representative of the
Virasoro orbit which leave it fixed under the coadjoint Virasoro action. The key observation
was noted in section \ref{sln} and is that the $(2,1)$ component of a gauge preserving
Kac-Moody transformation generates the conformal transformations, which means that for
determining the isotropy vectorfields one has to look for the $(2,1)$ component of
those infinitezimal Kac-Moody transformations which leave a particular gauge fixed current invariant.
A general Virasoro representative was obtained from a 
constant $J_0$ by applying first $T_n$ and then $h$ to bring it to the gauge fixed form, so
those elements which leave them invariant are obviously of the form
$h T_n A T_n^{-1} h^{-1} \in L\SLketto$, where $A \in \SLketto$ commutes with the monodromy
corresponding to $J_0$. Hence, the isotropy vectorfields are given by the $(2,1)$ component of
$h T_n Y T_n^{-1} h^{-1} \in L\slketto$, where
$Y = \left( \begin{array}{cc} y_1 & y_2-y_3 \\ y_2+y_3 & -y_1 \end{array} \right)$
is any element of the Lie algebra of the commutant subgroup of $M$. The $(2,1)$ component
turns out to be
$$V = \frac{y_3+y_2 \cos(2nx)+y_1 \sin(2nx)}{n+j_3+j_2 \cos(2nx)+j_1 \sin(2nx)} \; ,$$
which can be directly related to the results of \cite{rovid}. Note that in any case
when $L$ is non-constant and periodic with period $\frac{\pi}{n}$ a discrete $\Z_n$ subgroup must appear
in the isotropy subgroup corresponding to rigid discrete rotations of the circle.

The case-by-case analysis proceeds as follows:

\begin{itemize}

\item Elliptic orbits, $n$ is an integer here and $R(x)=\frac{1}{\sqrt{n+\omega}}$ where
$\omega \in (0,1)$ and $\omega \neq \frac{1}{2}$ so $n \geq 0$. The Virasoro representatives
and stabilizing vectorfield (up to an irrelevant normalization factor) are
$$L=-(n+\omega)^2 \; , \qquad V=1 \; .$$
The $V$ vectorfield generates an $S^1$.

\item Hyperbolic orbits, $n$ is a half-integer here and $R(x)=\frac{1}{\sqrt{n+b \sin(2nx)}}$
where $b > 0$. The expression under the square root can be negative for large $b$ but
if $J_0$ is conjugated by 
$\left( \begin{array}{cc}
1 & 0 \\
b & 1
\end{array} \right)$
then from the new $J_n$ one obtains $R(x)=\frac{1}{\sqrt{n+b^2+b^2 \cos(2nx)+b \sin(2nx)}}$
which is well defined, where $n \geq 0$. The Virasoro representatives and stabilizing vectorfield are
$$L=b^2+\frac{2n(n^2+b^2)}{n+b^2+b^2 \cos(2nx)+b \sin(2nx)}+\frac{3n^2(b^2-2b^2n-n^2)}{(n+b^2+b^2 \cos(2nx)+b \sin(2nx))^2} \; ,$$
and
$$V= \frac{b+\sin(2nx)+b\cos(2nx)}{n+b^2+b\sin(2nx)+b^2 \cos(2nx)} \; ,$$
which for $n=0$ generates an $S^1$ and for $n \neq 0$ a one-parameter subgroup isomorphic to $\R$.

\item Parabolic orbits, $n$ is a half-integer here and $R(x)=\frac{1}{\sqrt{n + \frac{q}{2\pi} \cos^2(nx)}}$. If $n=0$ only $q=1$ is allowed, otherwise $n>0$ and $q=\pm 1$ is arbitrary.
The Virasoro representatives and stabilizing vectorfields are
$$L=n^3 \left( \frac{2}{n + \frac{q}{4\pi} (1+\cos(2nx))} -
\frac{3(n + \frac{q}{2\pi})}{(n + \frac{q}{4\pi} (1+ \cos(2nx)))^2} \right) \, , \; V= \frac{1+\cos(2nx)}{n + \frac{q}{4\pi} (1+\cos(2nx))} \; ,$$
which again generates an $S^1$ for $n=0$ and a subgroup isomorphic to $\R$ for $n \neq 0$.
The above form of $L$ explicitly shows that if $n=0$ then only one of the $q$s are allowed
because if it were not so then two distinct monodromies would correspond to the same $L=0$.

\item Exceptional orbits, $n>0$ and is a half-integer here and $L$ can be obtained from
the elliptic case as the $\omega \nyil 0$ limit, but the stabilizing subgroup is larger. The
Virasoro representatives and stabilizing vectorfields are
$$L=-n^2 \; , \qquad V=1 , \; \sin(2nx) , \; \cos(2nx) \; .$$
These $V$s generate a $2n$-fold ``embedding'' of $\SLketto$ into Diff$(S^1)$.

\end{itemize}

It was noted in section \ref{sln} that the energy can have critical points only on those orbits
which contain a constant representative. To study the stability of the critical points
let us spell out the (\ref{ddenergy}) second variation allowing now only such $K$s
that preserve the gauge. Having a constant $L$ at hand the form of these on the
Lie algebra level are
$$K= \frac{1}{2} \left( \begin{array}{cc} \varepsilon^{\prime} & \varepsilon L - \varepsilon^{\prime \prime} \\ 2 \varepsilon & - \varepsilon^{\prime} \end{array} \right)$$
according to the infinitezimal version of formula (\ref{megorzo})
where $\varepsilon: S^1 \nyil \R$ is some function.
Calculation of the diagonal element of the (\ref{ddenergy}) second variation gives
$$\delta \delta E = 2 \int_0^{2\pi} (L {\varepsilon^{\prime}}^2 + \frac{1}{4} {\varepsilon^{\prime \prime}}^2) \frac{dx}{2\pi}$$
which is easily seen to be non-negative for any $\varepsilon$ if and only if $L \geq - \frac{1}{4}$.

\section{$\SLharom$ Kac-Moody and Zamolodchikov algebra}
\label{sl3w}

In the $\SLharom$ WZNW model two fields survive the reduction, the $W_2=L$ energy-momentum
tensor and the $W_3=W$ field. There are two
free functions in the infinitezimal version of the (\ref{megorzo}) gauge
preserving transformations out of which
the second component of $\mathbf a_1$ generates the conformal and the first component the
$W$-transformations. Applying such an infinitezimal transformation to a gauge fixed current
one can read off the Poisson brackets \cite{todatheory}:
\bea
\{L(x),L(y)\} & = & L^{\prime}(x) \delta(x-y) - 2L(x) \delta^{\prime}(x-y) - 2 \delta^{\prime \prime \prime}(x-y) \nn \\
\{L(x),W(y)\} & = & W^{\prime}(x) \delta(x-y) + 3 W(x) \delta^{\prime}(x-y) - \nn \\
& & - (L(x) \delta(x-y))^{\prime \prime} + \delta^{\prime \prime \prime \prime}(x-y) \nn \\
\{W(x),W(y)\} & = & \left( \frac{2}{3} L^{\prime}(x) L(x) + W^{\prime \prime}(x)
- \frac{2}{3} L^{\prime \prime \prime} \right) \delta(x-y) + \nn \\
& & + \left( \frac{2}{3} L^2(x) +2W^{\prime}(x)-2L^{\prime \prime}(x) \right) \delta^{\prime}(x-y)- \nn \\
& & - 2L^{\prime}(x) \delta^{\prime \prime}(x-y) - \frac{4}{3}L(x) \delta^{\prime \prime \prime}(x-y)+ \frac{2}{3} \delta^{\prime \prime \prime \prime \prime}(x-y) \nn
\eea
where the $W$ field is not primary but if gauge fixing
is changed from the Wronskian to the highest weight one \cite{todatheory, a2toda}, which is essentialy
a shift, $\widetilde{W}=W - \frac{1}{2} L^{\prime}$, then the Poisson brackets for the new fields
show that $\widetilde{W}$ is a primary
field with conformal weight $3$ and that its bracket with itself gives polinomial
terms in the energy-momentum tensor and happens to be identical to the Zamolodchikov algebra.

As the fundamental group of $\SLharom$ is $\Z_2$ each Kac-Moody orbit consists of
two connected components. The
\bea
\label{tn}
T_n(x)= \left( \begin{array}{ccc} \cos(nx) & -\sin(nx) & 0 \\ \sin(nx) & \cos(nx) & 0 \\ 0 & 0 & 1 \end{array} \right)
\eea
transformation determines the non-trivial element in $\pi_1(\SLharom)$ if $n$ is odd. This
means that for $n$ odd $T_n$ shifts between the two components, for $n$ even $T_n$
preserves them. This will result the interesting consequence that on some Kac-Moody orbits
and on the corresponding $W$-leaves there will be infinitely many critical points
of the energy but at most one will be a minimum.

The exploration of the $W$-leaves will proceed in a completely analogous way as in
the previous section. Since the fundamental group has two elements there corresponds
at least two symplectic leaves to each monodromy matrix but it is known from the context
of homotopy types of non-degenerate curves on projective space or sphere
that in some cases this classification is not complete as sometimes the constraint
intersects one of the two components of the Kac-Moody orbits in two pieces, which
will mean that to such a monodromy $3$ leaves correspond \cite{symplectic, shapiro}. 

A complete list of conjugacy classes with the isotropy subgroups follows. Note that the isotropy
subgroups have dimension $2$, $4$ or $8$ so an $(L, W)$ pair can be left fixed by only
$2$, $4$ or $8$ linearly independent $W$-transformations.

\begin{itemize}

\item  Hyperbolic classes, the diagonalizable matrices with $3$ distinct, real eigenvalues
belong here. 
$$H=\left\{ \left. \left( \begin{array}{ccc} t & 0 & 0 \\ 0 & r & 0 \\ 0 & 0 & (tr)^{-1} \end{array} \right) \right| \; \; t,r \neq 0 \; \right\} \simeq \R^* \times \R^* \; ,$$
representatives:
$$M= \left( \begin{array}{ccc} \pm e^{2\pi a} & 0 & 0 \\ 0 & \pm e^{2\pi b} & 0 \\ 0 & 0 & e^{-2\pi(a+b)} \end{array} \right) \; .$$

\item Degenerate hyperbolic classes, the diagonalizable matrices with $2$ distinct, real
eigenvalues belong here.
$$H=\left\{ \left. \left( \begin{array}{ccc} t & r & 0 \\ u & v & 0 \\ 0 & 0 & (tv-ru)^{-1} \end{array} \right) \right| \; \; tv-ru \neq 0 \; \right\} \simeq GL(2,\R) \; ,$$
representatives:
$$M= \left( \begin{array}{ccc} \pm e^{2\pi b} & 0 & 0 \\ 0 & \pm e^{2\pi b} & 0 \\ 0 & 0 & e^{-4\pi b} \end{array} \right) , \; b \neq 0 \; \; \mbox{in the ``+'' case} \; .$$

\item Elliptic classes, those matrices belong here which are diagonalizable over the complex
numbers with one real and a complex and its conjugate eigenvalue.
$$H=\left\{ \left. \left( \begin{array}{ccc} t & -r & 0 \\ r & t & 0 \\ 0 & 0 & (t^2+r^2)^{-1} \end{array} \right) \right| \; \; t^2+r^2 \neq 0 \; \right\} \simeq \C^* \; ,$$
representatives:
$$M= \left( \begin{array}{ccc} e^{2\pi b} \cos(2\pi \omega) & - e^{2\pi b} \sin(2\pi \omega) & 0 \\ e^{2\pi b} \sin(2\pi \omega)  & e^{2\pi b} \cos(2\pi \omega) & 0 \\ 0 & 0 & e^{-4\pi b} \end{array} \right) , \; \omega \in (0,1), \; \omega \neq \frac{1}{2} \; .$$

\item 1. parabolic classes, matrices with one degenerate Jordan block belong here. 
$$H=\left\{ \left. \left( \begin{array}{ccc} t & r & 0 \\ 0 & t & 0 \\ 0 & 0 & t^{-2} \end{array} \right) \right| \; \; t \neq 0 \; \right\} \simeq \R \times \R^* \; ,$$
representatives:
$$M= \left( \begin{array}{ccc} \pm e^{2\pi b} & 1 & 0 \\ 0 & \pm e^{2\pi b} & 0 \\ 0 & 0 & e^{-4\pi b} \end{array} \right) \; , b \neq 0 \; \; \mbox{in the ``+'' case} \; .$$

\item 2. parabolic class.
$$H=\left\{ \left. \left( \begin{array}{ccc} t & r & u \\ 0 & t & 0 \\ 0 & v & t^{-2} \end{array} \right) \right| \; \; t \neq 0 \; \right\} \simeq \R^3 \times \R^* \; ,$$
representative:
$$M= \left( \begin{array}{ccc} 1 & 1 & 0 \\ 0 & 1 & 0 \\ 0 & 0 & 1 \end{array} \right) \; .$$

\item 3. parabolic class.
$$H=\left\{ \left. \left( \begin{array}{ccc} 1 & t & r \\ 0 & 1 & t \\ 0 & 0 & 1 \end{array} \right) \right| \; \; t,r \in \R \right\} \simeq \R^2 \; ,$$
representative:
$$M= \left( \begin{array}{ccc} 1 & 1 & 0 \\ 0 & 1 & 1 \\ 0 & 0 & 1 \end{array} \right) \; .$$

\item Exceptional class.
$$H=\SLharom , \; \mbox{representative:} \; M=1 \; .$$

\end{itemize}

At the hyperbolic, degenerate hyperbolic and the 1. parabolic class the signs are
either both positive or both negative.

The classification of $W$-leaves by the fundamental group is complete in the ``$-$'' version of the hyperbolic, the degenerate
hyperbolic,  and the 2. parabolic case. In the others the constraint intersects one of the two 
Kac-Moody components in two piece, more concretely in the exceptional case $n=1$ corresponds
to a separate leaf, while all other odd $n$ correspond to a second, all even $n$ to a third
leaf. In the remaining cases $n=0$ constitutes to a separate leaf, all other even $n$ to a second
and all odd $n$ to a third leaf \cite{symplectic, shapiro}.

The given representatives possess logarithms except for the 1. parabolic and the ``$-$''
version of the hyperbolic classes. The latter case was analyzed in \cite{a2toda} and
the authors did not find constant $L$ and $W$ representatives for the corresponding
leaf which is not surprising as we have noted in section \ref{hurok} that if a
monodromy does not have a logarithm then the corresponding orbit can not have a constant
representative.

In the ``$-$'' cases the trick used for $\SLketto$ will be adopted in the upper left
block, i.e. take the opposite sign of the upper left $2 \times 2$ block of $M$, call 
it $M_-$ which possess a logarithm and then the monodromy of
$g(x)=T_{\frac{1}{2}}(x)M_{-}^{\frac{x}{2\pi}}$ will be $M$ where $T_{\frac{1}{2}}(x)$ is
the matrix of rotation by $\frac{x}{2}$ in the upper left block. On the resulting
matrices the $T_n$ matrix must be applied but note again that for $n$ even we will stay
in the original component. The resulting matrices should be transformed into the gauge
fixed form by a Kac-Moody transformation that determines the trivial element of $\Z_2$.

Let us analyze the stability of the critical points on the $W$-leaves which is
important for determining the classical h. w. states. Substituting
the gauge preserving infinitezimal transformations into the (\ref{ddenergy}) diagonal
element of the second variation of the energy gives:
$$\delta \delta E =\int_0^{2\pi} \left( \begin{array}{cc} \varepsilon_1^{\prime} & \varepsilon_2^{\prime} \end{array} \right) \left( \begin{array}{cc} \frac{2}{3} \left( L- \frac{d^2}{dx^2} \right)^2 & 3W-\left( L- \frac{d^2}{dx^2} \right) \frac{d}{dx} \\ 3W+\left( L- \frac{d^2}{dx^2} \right) \frac{d}{dx} & 2 \left( L - \frac{d^2}{dx^2} \right) \end{array} \right) \left( \begin{array}{c} \varepsilon_1^{\prime} \\ \varepsilon_2^{\prime} \end{array} \right) \frac{dx}{2\pi}$$
supposed $L$ and $W$ are constant, where $\varepsilon_1$ and $\varepsilon_2$ are the functions generating
the conformal and $W$-transformations. This quadratic form is non-negative for any
$\varepsilon_1$ and $\varepsilon_2$ if and only if
$$(L+1)^2 (4L+1) \geq 27 W^2$$
holds for the constant representatives, which shows that $L \geq -\frac{1}{4}$ is necessary just as in
the pure Virasoro case.

The straightforward calculation gives in the individual cases the
following, where we have not bothered writing out explicitly the non-constant representatives
as they are quite complicated and have no transparent physical meaning.

\begin{itemize}
\item Hyperbolic orbits. In the $n=0$ and ``$+$'' case one has constant representatives
which can be obtained by constant Kac-Moody transformations:
$$L=a^2+ab+b^2, \quad W=-ab(a+b)$$
which are all minima of the energy. If $n \neq 0$ there are only non-constant $L$ and $W$ 
representatives.

\item Degenerate hyperbolic orbits. In this case $n$ is a half-integer, the classification is complete by the
fundamental group and we have constant representatives for the $W$-leaves:
$$L=3b^2-n^2, \quad W=-2b(b^2+n^2) \; .$$
Note again that these elements describe $4$ different leaves, according to whether
$n \neq 0$ is an integer or half-integer (the monodromies are different) and whether
$n$ is even or odd in case it is an integer, or $n-\frac{1}{2}$ is even or odd in
case $n$ is a half-integer (same monodromy but different Kac-Moody component).
Now any leaf possesses infinitely many critical points but only in the ``$-$'' case
will one find a minimum and it only happens if $n= \frac{1}{2}$ and $b$ arbitrary and the minimum
is degenerate. Note that these classical h. w. states are $GL(2,\R)$ invariant so
they are good candidates for a true vacuum in CFT.

\item Elliptic orbits. Here $n$ is an integer and one obtains constant representatives
for any $W$-leaf:
\bea
\label{lw}
L=3b^2-(\omega+n)^2 , \; \; W=-2b(b^2+(\omega+n)^2), \; \; \omega \in (0,1),\; \omega \neq \frac{1}{2} \; .
\eea
In these cases the classification by the fundamental group is not complete, the constraint intersects
the component corresponding to even $n$ in two piece, but the $W$-leaf corresponding
to odd $n$ is connected and the (\ref{lw}) representatives determine the same $W$-leaf
for all odd $n$ which means that on these leaves the energy has infinitely many critical
points. The critical points are minima only if $n=0$ and $|\omega| < \frac{1}{2}$ and $b$ arbitrary.

\item 1. parabolic orbits. These have constant representatives for $n=0$ and they are
given by
$$L=3b^2, \quad W=-2b^3 \; ,$$
which are minima of the energy for arbitrary $b$.

\item 2. parabolic orbits. None of these orbits have constant representatives.

\item 3. parabolic orbit. The constant representative is
$$L=0, \quad W=0 ,$$
which is a minimum of the energy.

\item Exceptional orbit. The leaf corresponding to this orbit has infinitely many constant
representatives,
$$L=-n^2, \quad W=0 \; ,$$
where $n \neq 0$ is an integer, but none of these are minima of the energy.

\end{itemize}

Note again that a constant representative with non-negative second variation of 
the energy is a classical h. w. state. One hopes to construct a consistent highest weight
representation space on these leaves on which the quantized version of the $W$-algebra
will act in CFT. The uniqueness of the vacuum in CFT is an important issue and
we have established that if there is a critical point of the energy on a leaf then at most
one will give rise to a vacuum as at most one of the possibly many critical points can
be a stable minimum. Those leaves which do not contain critical points, that means on which
one can not find a constant representative, are physically less important as 
one can not associate a heighest weight type representation of the $W$-algebra to these. 

\section{Conclusions and outlook}
\label{conc}

We have analyzed in this paper the classical geometry of $W$-algebras in the context of
reduced $\SLn$ WZNW models. We split the WZNW field as 
$$g(x,t)=g_{+}(x_{+})g_{-}(x_{-})$$
and concentrated on one chiral half of the original WZNW theory only. As a
consequence of the splitting the chiral theory inherited a global $G$-symmetry
$ g_{+}\to g_{+}h\, ,\, \, h\in G$. Both the unconstrained and constrained currents are invariant under this
global symmetry, since
$$J_{+}=(\partial _{+}g) g^{-1} \, ,$$
that is the fields $W_{2}, \dots , W_{n}$
are $\SLn$ invariant. We fixed this gauge symmetry, however, by demanding
$g(0)=e$ and establishing a one to one correspondence between the current
$J_{+}$ and the group valued field $g_{+}$. In the reduced case this
means that the fields $W_{2}, \dots ,W_{n}$ can be determined from the last
row of $g$ , that is from a non-degenerate curve in $\R^{n}$ as we have
seen. If we had not fixed the gauge we would have the global $\SLn$ gauge
symmetry. The geometry of this symmetry was discussed in detail in \cite{affine}.
Let's summarize their results. This symmetry acts naturally on the space of non-degenerate
curves and the generators of the classical W-algebra, that is the components
of the constrained currents $W_{2}, \dots ,W_{n}$ are invariant. Since they can
be built up from the non-degenerate curve as differential
polynomials which are clearly invariant with respect to the gauge symmetry they
can be interpreted as ``affine'' invariants. It was also shown that they
are the only invariants and determine the curve modulo gauge transformations.
The Virasoro generator is a second order invariant which is called ``affine''
curvature, the next $W$-generator can be interpreted as ``affine'' torsion, etc.

Here we have concentrated on the classification of $W$-leaves and the issue of
classical h. w. states in one chiral half. We can extend our results
for the other chiral half of the theory since analogous formulae hold for them.
Putting the two chiral halfs together the solutions of the reduced WZNW
models, namely the Toda models, can be obtained. Our classification of the leaves
leads to a classfication of the $W$-inequivalent Toda solutions as it was
described in detail in \cite{coadjoint1} for the global Liouville equation.
Our method in principle is able to give explicit representatives
on each leaf which fact was put in practice in the search for classical h. w. type
representations. All these considerations were carried out
for $n=2$ and $n=3$ which corresponds to the Virasoro and Zamolodchikov algebra respectively.
We have found that there is at most one classical h. w. state on any $W$-leaf
and in the Zamolodchikov algebra case also determined the $\SLketto$ ($GL(2,\R)$ in fact)
invariant highest weight states which are the best candidates for a true vacuum in CFT.

There is a very interesting and delicate question, how the classical representions
are related to the quantum highest weight representations.
One possible approach is to take the classical equations of motion and try to
quantize them. This idea was used in \cite{a2toda, locality, c2toda} and they found
that consistent quantization results in $W$-minimal models, where the classical
Toda field corresponds to the simplest nontrivial completely degenerate representation
of the quantum $W$-algebra. Later it was extended by taking higher dimensional
representions of the reduced WZNW model to obtain other representions in the
quantum case in \cite{newapp}. It is not clear however how this representions are
related to the symplectic leaves of the classical $W$-algebra even in the simplest
Virasoro case, where only a conjecture from Witten gives some hint \cite{coadjoint2}. A proper description
would be the geometric quantization of the leaves classified in this paper. This is
a difficult problem and is unsolved even in the finite $W$-algebra case except for
the simplest situation \cite{sajat}.

\section*{Acknowledgements}
The work of Z. B. was supported by the grants OTKA D25517 and T029802.

\end{document}